\documentclass{iau} 
\usepackage{graphicx}

\title[Radial Distribution of Blue Stragglers in NGC 5466 using AstroSat] 
{Radial Distribution of Blue Stragglers in NGC 5466 using AstroSat}

\author[Snehalata Sahu \& Annapurni Subramaniam]   
{Snehalata Sahu$^{1,2}$
 \and Annapurni Subramaniam$^1$}

\affiliation{$^1$Indian Institute of Astrophysics\\ Koramangala II Block, Bengaluru, 560034, India \\ 

$^2$Pondicherry University\\ R.V. Nagar, Kalapet, 605014, Puducherry, India \\
email: {\tt snehalata@iiap.res.in, purni@iiap.res.in}}

\pubyear{2019}
\volume{351}  
\jname{Star Clusters: From the Milky Way to the Early Universe}
\editors{A. Bragaglia, M.B. Davies, A. Sills \& E. Vesperini, eds.}
\begin{document}

\maketitle

\begin{abstract}
We present the results obtained from Near-UV observations of the cluster NGC 5466 taken with UVIT onboard AstroSat to study the radial distribution of Blue Straggler Stars (BSSs), covering the cluster region up to a radius of 14$'$. Our study confirms that the BSSs are more centrally concentrated than Horizontal Branch (HB) stars in the cluster. We do not find a statistically significant rising trend in the radial distribution of BSSs for the outer regions, as most of the previously catalogued BSSs that are located in the cluster outskirts were found to be non-members, quasars or galaxies. This study highlights the importance of UV imaging combined with membership information to probe the radial distribution of BSSs. 
\keywords{(Galaxy:) globular clusters: individual (NGC 5466), (stars:)- Blue Stragglers, stars: horizontal-branch, ultraviolet: stars.}
\end{abstract}

\section{Introduction}
NGC 5466 is an old, metal poor ($[Fe/H]=-2.0$, \cite[Caretta \etal\ 2009]{Care_etal2009}) Globular Cluster (GC) located in the constellation Bo\"{o}tes. It is a low density GC (log$_{10}$ $\rho_{c} \sim$ 0.84 $L_{\odot}/pc^{3}$, \cite[McLaughlin \& Van der Marel 2005]{Mcvan_2005}) as compared to other Galactic GCs of similar luminosity.
\cite[Nemec \& Harris (1987)]{Nem_har1987} first studied and analysed the Blue Straggler Star (BSS) population covering a radius $\sim 9.1'$ from the cluster center using Canada-France-Hawaii Telescope (CFHT). They identified 48 BSSs from optical Color-Magnitude Diagram (CMD). Later, \cite[Fekadu \etal\ (2007)]{fek_etal2007} provided a catalog of 94 BSS candidates and studied the radial distribution of the BSS population till r $\sim13'$ from the cluster center. Recently, \cite[Beccari \etal\ (2013)]{Becca_etal2013} (B13 hereafter) combined the data from Advanced Camera for Surveys (ACS), LBC, and CFHT data to study the BSS population up to the tidal radius of the cluster (r $\sim$ 26.3$'$). They found that the cluster contains a large fraction of binaries ($\sim 6\%$) and BSSs. They also concluded that the cluster has a bimodal radial distribution. 

\cite[Mateo \etal\ (1990)]{Mat_etal1990} discovered 2 contact binaries and 1 eclipsing binary suggesting that the evolution of primordial binaries in isolation in such a low density environment would have led to the BSS formation. Recently, \cite[Sahu \etal\ (2019)]{sahu_2019} identified a hot companion to a BSS located in the cluster outskirts using the far-UV (FUV) and near-UV (NUV) observations obtained by the UltraViolet Imaging Telescope (UVIT), thus, showing the advantage of UV photometry in identifying such BSS binary candidates. 

Previous studies on radial distribution of BSSs were mainly based on the BSS selection from the optical CMDs. The cluster field is contaminated by a large number of background sources which can cause a problem in the selection of BSSs using optical CMDs. In this study, we have addressed these issues by considering the following: 1) selecting BSSs using UV CMDs i.e. by switching to UV wavelengths instead of optical and, 2) determining the proper motion (PM) membership of the BSSs which is extremely important in the outer regions of the cluster where the chance of false detection is very high. For achieving the first aim, we have used images from UVIT onboard AstroSat, the first Indian space observatory and combined it with the HST-ACS data by \cite[Sarajedini \etal\ (2007)]{Sara_2007} and ground data provided by P. Stetson for the regions outside the field of view of HST. To achieve the second aim, we have used Gaia DR2 data (\cite[Gaia Collaboration \etal\ 2018]{Gaia_2018}) to select the BSS members. The UVIT image, observations and photometry details of the cluster are given in \cite[Sahu \etal\ (2019)]{sahu_2019}.

\begin{figure}
\begin{center}
\includegraphics[width=\columnwidth]{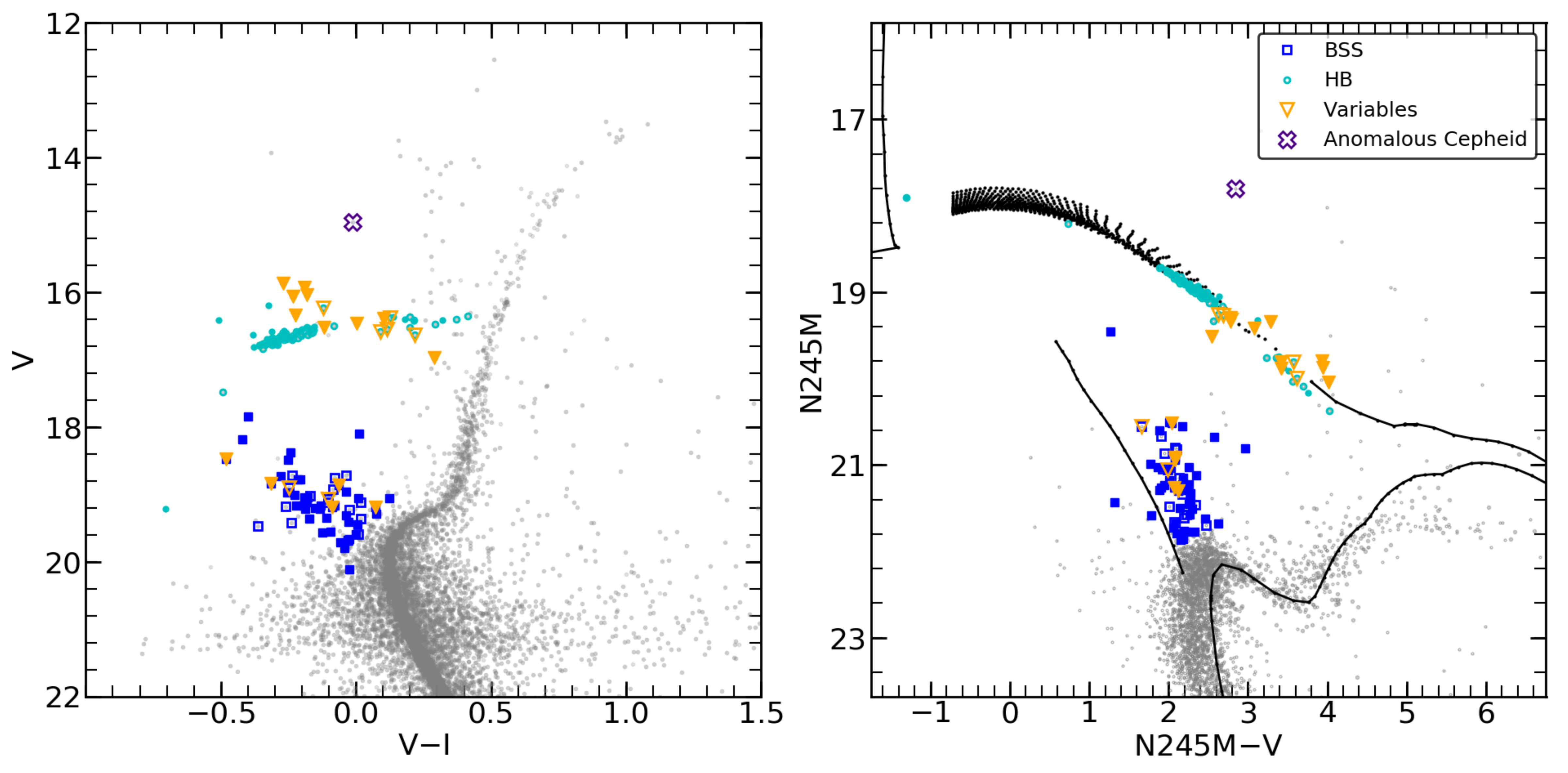}
\caption{\textit{Left:} V$-$I vs V optical CMD \textit{Right:} N245M$-$V vs N245M CMD where the closed symbols are UVIT cross-matched HST detections and open symbols are UVIT cross-matched ground detections. The different symbols marked in the legend are: BSSs (blue squares), HB stars (cyan circles) and the known Variables including RR Lyrae and SX Phes (orange triangles).}
\label{nuvcmd}
\end{center}
\end{figure}

\section{NUV and optical CMDs}
The NUV (N245M filter of UVIT) data was cross-matched with optical data from HST-ACS and ground data to construct the NUV-optical CMD covering the full cluster region (r$\sim$14$'$). The data was then cross-matched with Gaia DR2 catalog (\cite[Gaia Collaboration \etal\ 2018]{Gaia_2018}) to select the PM members. The PM cleaned N245M$-$V vs N245M CMD is shown in the right panel of Figure \ref{nuvcmd} along with its corresponding optical CMD in the left panel. The black line and dots correspond to a Padova isochrone of [Fe/H]= $-$1.98 dex and age 12.6 Gyr generated using FSPS models \cite[(Conroy \etal\ 2009)]{conroy_2009}. We cross-matched the BSS catalog provided by \cite[Fekadu \etal\ (2007)]{fek_etal2007} with the Gaia DR2 catalog to identify the PM members. We found 8 of them to be non members among which 3 are classified as quasars by \cite[Flesch (2015)]{Flesch2015}. We also found 3 of the candidates which do not have PM, to be classified as galaxies by the SDSS surveys. These contaminants are mostly located in a region outside the half-right radius ($r_{h} \sim 140''$, \cite[McLaughlin \& Van der Marel 2005]{Mcvan_2005}) of the cluster. We plotted them in the N245M vs N245M$-$V CMD to check their locations and rejected 15 of the BSS candidates from our sample as they do not qualify our criteria of finding them in the expected BSS location in the CMD. They were either found in the SGB location or they had redder NUV colors (N245M$-$V $>$ 3.0). We are left with a final sample of 66 BSSs (7 SX Phes and 3 binaries) which are PM members and we use this sample to study the radial distribution of BSSs in the cluster.

\subsection{Radial distribution of BSSs}
\begin{table}
\centering
\caption{Number of BSSs and HB stars in the radial annuli that are considered to study the radial distribution of BSSs. R$_{in}$ and R$_{out}$ are the inner and outer radius of the radial bin selected for counting the BSSs and HB stars. Column 5 gives the number of BSSs detected by B13.}
\begin{tabular}{ccccc}
\hline
$R_{in}^{''}$ & $R_{out}^{''}$ & $N_{BSS}$ & $N_{HB}$ & $N_{BSS}^{Beccari}$ \\\hline
0	&	72	&	31	&	24	&	32\\
72	&	150	&	25	&	18	&	37\\
150	&	200	&	4	&	12	&	4\\
200	&	500	&	5	&	28	&	11\\
500	&	850	&	1	&	4	&	10*\\\hline
\multicolumn{5}{c}{*for $R_{out}^{''}$ = 1400 (B13)}
\end{tabular}
\label{rad_bss}
\end{table}

\begin{figure}
\begin{center}
\includegraphics[width=0.47\columnwidth]{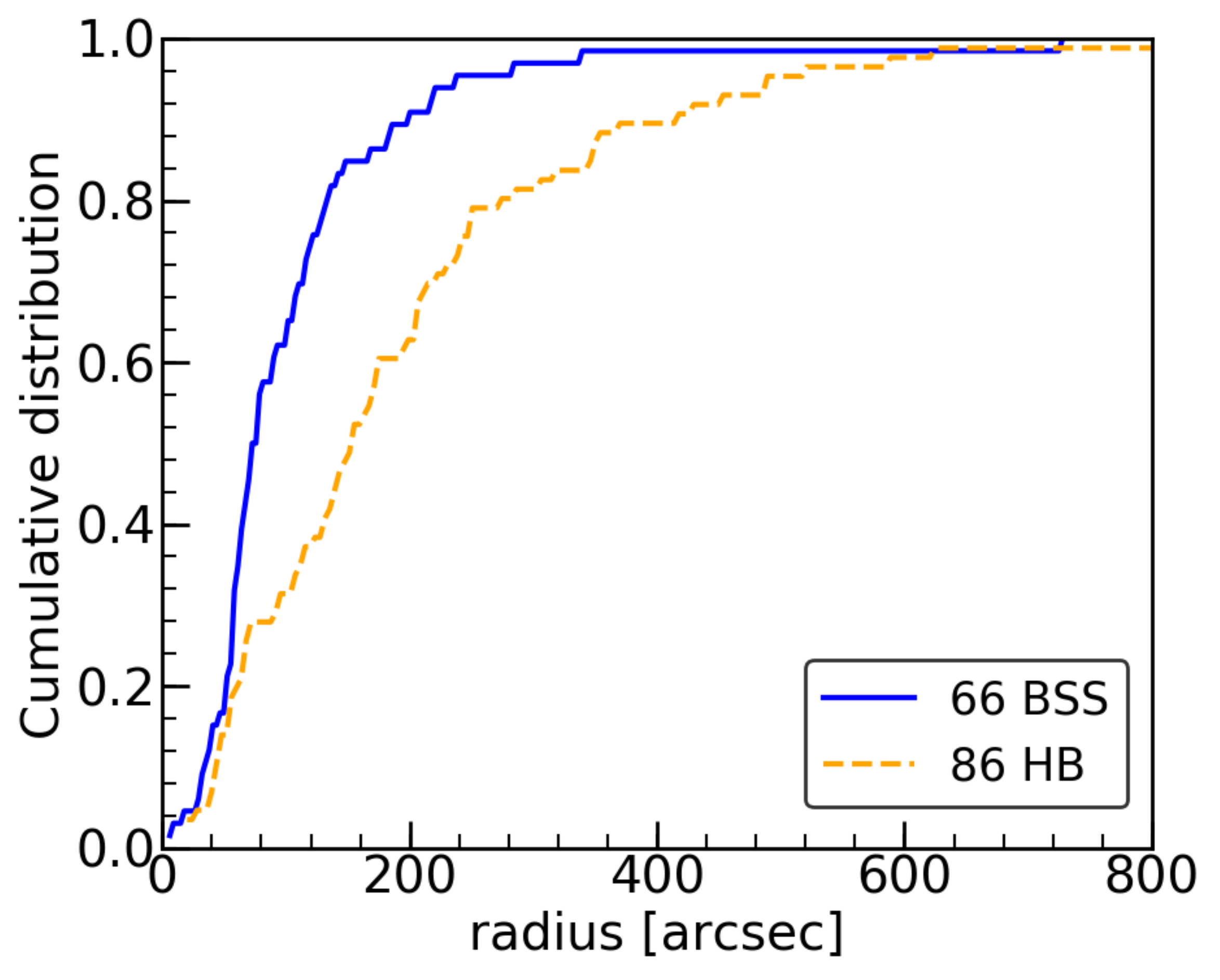}
\includegraphics[width=0.48\columnwidth]{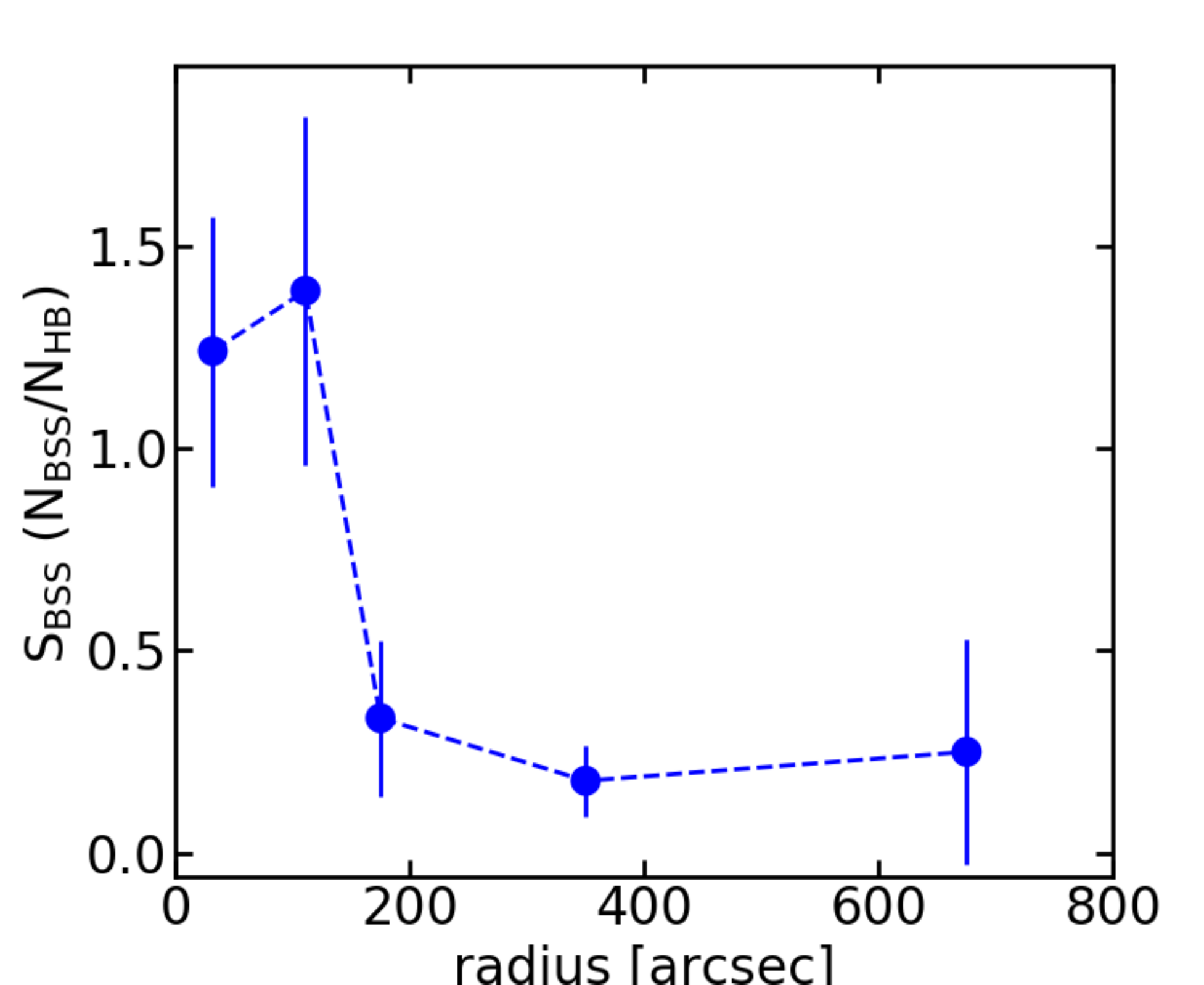}
\caption{\textit{Left:} Cumulative radial distribution of BSSs (blue line) and HB stars (orange dashed line) and \textit{Right:} Specific frequency of BSSs (S$_{BSS}$) as a function of radial distance from the cluster center.}
\label{cumbss}
\end{center}
\end{figure}

We selected 86 HB stars that are PM members as a reference population for studying the radial distribution. The cumulative radial distribution of BSSs and HB stars are shown in the left panel of Figure \ref{cumbss}. We can clearly notice that the BSSs are more centrally concentrated than HB stars. According to the Kolmogorov-Smirnov test, the probability that BSSs and HB stars are drawn from the same population is negligible with p-value $\sim 5.7 \times 10^{-6}$. 

We derived the specific frequency of BSSs (S$_{BSS}$) which is defined as the ratio of number of BSSs (N$_{BSS}$) to the number of HB stars (N$_{HB}$). To do this, we divided the observed cluster region into several concentric annuli and counted the number of BSSs and HB stars in each radial bin which is given in Table \ref{rad_bss}. The radial annuli selected for this study are similar to those given by B13 except in the outermost annulus where the outer radius ($R_{out}$) of the radial bin is 850$''$ instead of 1400$''$. Besides, they used red giant stars as reference population instead of HB stars. The comparison of the number of BSSs found in each radial bin using our sample with those reported by B13 is also given in Table \ref{rad_bss}. The specific frequency of BSSs as a function of radial distance is shown in the right panel of Figure \ref{cumbss}. We find that the distribution shows a first dip or minimum at $r_{min} \sim 180''$ which is in agreement with B13. This minimum is close to 1.5$r_{h}$ of the cluster suggesting that the dynamical friction has affected the BSSs located in the internal regions ($< 1.5r_{h}$) causing them to segregate towards the cluster center. According to the study by B13, there was a rising trend in the distribution at $r< 500''$ unlike, the decreasing trend that we found in our study. This is because the number of BSSs in their sample with radial interval $200'' < r < 500''$ was 11 whereas it reduces to 5 in our sample after considering the PM selection. We do not find any upturn in the distribution at radial interval  $500 < r < 850''$ which is subjected to large error due to poor statistics. This large error in the estimation along with the incomplete coverage (till radius $\sim$ 14$'$) prevents us from confirming whether the rise in the upturn as reported in the earlier study is significant or not. Further, the presence of background contaminants suggests that extreme care has to be taken while selecting the BSSs that are located in the outer regions of the cluster.

\section{Conclusion}
The r$_{min}$ in the radial distribution confirms the result by \cite[Beccari \etal\ (2013)]{Becca_etal2013} that NGC 5466 is the second youngest in the category of dynamically intermediate-age clusters (Family II, as defined by \cite[Ferraro \etal\ 2012]{ferr_2012}). This implies that the innermost BSSs have already segregated towards the cluster center unlike, the outer ones which have not yet experienced the mass segregation and are evolving in isolation. In such cases, we should expect a bimodal radial distribution which is not found in our results. \cite[Fekadu \etal\ (2007)]{fek_etal2007} pointed out that the tidal stripping of the cluster could be responsible for causing the removal of BSSs from the cluster potential, that are located in the cluster outskirts. Therefore, we find a dearth of BSSs in the outer regions resulting in a single radial distribution. This study thus demonstrates the significance of using UV photometry and membership information to study the BSS population. 

\section*{Acknowledgements}
This publication uses the data from the AstroSat mission of the Indian Space Research Organisation (ISRO), archived at the Indian Space Science Data Centre (ISSDC) which is a result of the collaboration between IIA, Bengaluru, IUCAA, Pune, TIFR, Mumbai, several centres of ISRO, and CSA.

\end{document}